\documentclass[aps,twocolumn,noshowkeys]{revtex4-2}
\usepackage{amsmath}
\usepackage{graphicx}
\usepackage{bm}
\usepackage{hyperref}
\usepackage{xcolor}
\usepackage{multirow}
\usepackage{amssymb}
\usepackage{soul}

\newcommand{\Eref}[1]{Eq.~(\ref{#1})}
\newcommand{\tref}[1]{Table~\ref{#1}}

\newcommand{\RaxeN}{R_{\textrm{ax}}^{(eN)}}
\newcommand{\Raxee}{R_{\textrm{ax}}^{(ee)}}

\begin{document}
\title{Static electric dipole moment of the francium atom induced by axionlike particle exchange}

\begin{abstract}
    The francium atom is considered as a prospective candidate system to search for the $\mathcal{T,P}$-violating electron electric dipole moment [Aoki \textit{et al.}  Quantum Sci. Technol. \textbf{6}, 044008 (2021)]. We demonstrate that the same experiment can be used for axionlike particles (ALP) search. For this, we calculate electronic structure constants of the ALP-mediated interaction for a wide range of ALP masses. Using the recently updated constraints on products of ALP-electron and ALP-nucleon coupling constants, we show that contribution of considered interactions corresponding to these constraints can give significant contribution to the atomic electric dipole moment. Therefore, obtainment of stronger restrictions for ALP characteristics in the francium atom electric dipole moment experiment is possible.
\end{abstract}

\author{D.E. Maison$^{1,2,*}$, 
L.V.\ Skripnikov$^{1,2}$}
\affiliation{$^{1}$Petersburg Nuclear Physics Institute named by B.P.\ Konstantinov of National Research Center ``Kurchatov Institute'' (NRC ``Kurchatov Institute'' - PNPI), 1 Orlova roscha, Gatchina, 188300 Leningrad Region, Russia}
\affiliation{$^{2}$Saint Petersburg State University, 7/9 Universitetskaya Naberezhnaya, St. Petersburg, 199034 Russia}
\homepage{http://www.qchem.pnpi.spb.ru}

\email{maison$\_$de@pnpi.nrcki.ru}

\maketitle

\section{Introduction}

Numerous experiments in modern particle physics are aimed at observation of the phenomena beyond the standard model (SM)~\cite{Safronova:18}. The development of the SM in the second half of twentieth century was an essential milestone in the  foundation of fundamental physical interactions theory. Observation of the Higgs boson confirmed the main ideas of the SM. However, at present some effects cannot be explained within the SM. The most popular example is the unknown nature of cosmological dark matter and dark energy. Moreover, SM is not able to explain the gravity mechanism and therefore surely needs an extension.

One of the approaches to search for beyond-SM effects is the investigation of $\mathcal{C}$-, $\mathcal{P}$- or $\mathcal{T}$-violating interactions \cite{Khriplovich:97, Ginges:04, Pospelov:05, fukuyama2012searching, nir1999cp, Barr:93b, ellis1982cp, branco2012leptonic, donoghue1978t}.  Here $\mathcal{C}$, $\mathcal{P}$ and $\mathcal{T}$ are the operations of the charge conjugation, spatial inversion and time reversion, respectively. In the 1960s, the study of the $K$-meson decay revealed that combined $\mathcal{CP}$ invariance is violated. According to the $\mathcal{CPT}$ theorem \cite{weinberg1995quantum},  $\mathcal{CP}$ violation implies also $\mathcal{T}$ violation, because the combined $\mathcal{CPT}$ symmetry must conserve. Current status of the  experiments to test $\mathcal{CPT}$ theorem can be found, i.e in Ref.~\cite{Safronova:18}. It was shown that $\mathcal{CP}$ violation is one of three necessary conditions to explain the observed predominance of the matter over the antimatter in the Universe \cite{sakharov1967violation}. 

The consequence of a simultaneous $\mathcal{T}$- and $\mathcal{P}$-invariance violation in particle physics is the existence of nonzero electric dipole moments (EDMs) of elementary particles. For example, numerous experiments with atoms \cite{murthy1989new, Regan:02, Graner:2016, rosenberry2001atomic} and molecules \cite{Hudson:02, Cornell:2017, ACME:14a, ACME:18} established strong constraints on the electron EDM ($e$EDM) value, $d_e$. The most precise one was obtained in the second generation of the ACME Collaboration experiment \cite{ACME:18}: $|d_e| \leq 1.1 \times 10^{-29}\ |e|\ \textrm{cm}$, where $e < 0$ is the electron charge.  However, this constraint is many orders of magnitude higher than the SM prediction~\cite{hoogeveen1990standard,Khriplovich:11,Yamaguchi:2021}.
Recent constraint on the neutron EDM is: $|d_n| \leq 1.8 \times 10^{-26}\ |e| \ \textrm{cm}$ \cite{abel2020_neutronEDM}. It is almost two times better than the previous one \cite{nEDM}.  The neutron EDM is of great interest since it provides information about the $\mathcal{CP}$ violation in the quantum chromodynamics sector~\cite{balachandran2012electric, gutsche2017c, zhevlakov2019bounds, zhevlakov2019updated,zhevlakov2020deuteron}. Besides, the proximity of $d_n$ to zero (so-called ``fine-tuning'')  was supposed not to be the coincidence, but the consequence of the spontaneous symmetry breaking \cite{peccei1977cp}. This symmetry was called Peccei-Quinn symmetry $U_{\textrm{PQ}}(1)$ and its breaking demands the existence of the pseudo-Goldstone boson, named afterwards \textit{axion} \cite{weinberg1978new, wilczek1978problem}.

Later, it was noted that axions and axionlike particles (ALP) are perfect candidates to be the dark matter component \cite{preskill1983cosmology, abbott1983cosmological,dine1983not}. Despite numerous experiments to search for ALPs \cite{Hare:2020, youdin1996limits, ni1999search, Duffy:2006, Zavattini:2006, hammond2007new, hoedl2011improved, barth2013cast, pugnat2014search, Abel:2017, flambaum2018resonant,aybas2021, Roussy:2021}, it has not been reliably detected. As it was noted in Refs.~\cite{graham2011axion,graham2013new}, the interaction of the cosmic ALPs with fermions can lead to oscillating atomic and molecular EDMs, which can be observed. In Ref. \cite{budker2014proposal} it was proposed to search for the varying EDM employing solid-state nuclear magnetic resonance technique. The results of this experiment are reported in Ref. \cite{aybas2021} (see also Ref.~\cite{Skripnikov:16a} for the theoretical study of the Schiff moment enhacement in crystals). Significant constraints on the axion-gluon coupling constants were also achieved in the experiment with the HfF$^+$ molecular cation~\cite{Roussy:2021}.

In Ref.~\cite{Stadnik:2018}, it was shown that the exchange by the virtual ALP between electron and nucleon or electron and electron can also lead to appearance of nonzero \textit{static} $\mathcal{T}$,$\mathcal{P}$-violating EDM of an atom or a molecule. Therefore, results of planned \cite{augenbraun2020laser, augenbraun2021observation} and already performed \cite{murthy1989new,Regan:02,Graner:2016,ACME:18,Cornell:2017,Hudson:11a} atomic and molecular experiments aimed to search for $e$EDM and other $\mathcal{T}$,$\mathcal{P}$-violating interactions can be also used to constrain ALP mass and products of ALP-fermion coupling constants. Due to arising interest to the YbOH molecule \cite{kozyryev2017precision, Hutzler:2020, Maison:2019b, denis2019enhancement, denis2020enhanced, zakharova2021rovibrational}, it  has been extensively studied from this point of view in Refs. \cite{Maison:2021, maison2021axion}.

It was suggested to perform the $e$EDM search experiment with the francium atomic beam~\cite{kawamuraFrancium:2013, Kawamura:14, inoueFrancium:2015, Harada:2016}. The scheme of the measurement using laser-cooled atoms trapped in optical lattice was suggested in Ref.~\cite{aoki2021quantum}. 
The sensitivity of such experiment is expected to be below $d_a(\textrm{Fr}) \lesssim 10^{-27}\ |e| \ \textrm{cm}$~\cite{aoki2021quantum}.

The sensitivity to $e$EDM in atoms with similar electronic structure is scaled roughly as $Z^3$~\cite{Sandars:65,Khriplovich:91}, where $Z$ is the nuclear charge. Moreover, most of currently considered $\mathcal{T}$,$\mathcal{P}$-violating effects are described by operators, whose mean values are mostly determined by the behavior of the valence wave function in the vicinity of the nucleus and determined mainly by $s_{1/2}$ and $p_{1/2}$ electronic states. For these reasons, $_{\ 87}^{210}$Fr atom with an open $7s$ electronic shell is one of the most prospective atomic candidates to measure $e$EDM. Theoretical studies of $\mathcal{T}$,$\mathcal{P}$-violating effects in the francium atom, such as $e$EDM and the scalar-pseudoscalar nucleus-electron interaction, have been performed in several works
\cite{Byrnes:99, Mukherjee:09, Blundell:2012, Roberts:2014, Skripnikov:17a, shitara2021cp} (see also references therein).

The present paper is devoted to  the study of the $\mathcal{T}$,$\mathcal{P}$-violating ALP-mediated nucleon-electron and electron-electron interactions in the francium atom. We perform precise calculation of electronic structure properties with various values of ALP masses and conclude that the expected sensitivity of the experiment~\cite{aoki2021quantum} with the francium atom can be sufficient for obtainment of the new constraints for the products of axion-fermion coupling constants.

\section{Theory}
The interaction of ALP with SM fermions is given by the Lagrangian,
\begin{equation} \label{Lagrangian}
    \mathcal{L} = a \sum\limits_\psi \bar{\psi}
    \left( g_\psi^s + g_\psi^p \gamma^5\right) \psi.
\end{equation}
Here $a$ is the axion field, $g_\psi^s$ and $g_\psi^p$ are scalar and pseudoscalar axion-fermion coupling constants, respectively, $\psi$ is the fermion field, $\bar{\psi} = \psi^\dagger \gamma^0$ and $\gamma^0$ and $\gamma^5$ are Dirac matrices defined according to Ref. \cite{Khriplovich:91}. The sum is over all fermions. QCD origin of the axion leads to relations between $g_{\psi}$ constants and axion mass $m_a$ \cite{holdom1982raising, di2020landscape}. However, for ALPs one usually assumes an independence of coupling constants and $m_a$. Below we follow this widely used agreement and do not distinguish QCD axions and ALPs and call them just axions.

The $\mathcal{T}$,$\mathcal{P}$-violating electron-nucleon interaction mediated by the axion is described by the following operator~\cite{moody1984new,gharibnejad2015dark,Stadnik:2018}:
\begin{equation} \label{intHam_eN}
    V_{eN}(\boldsymbol{r}) = i \frac{g_N^s g_e^p}{4\pi} \frac{e^{-m_a r}}{r} \gamma^0 \gamma^5.
\end{equation}
Here $\boldsymbol{r}$ is the radius vector of the electron with respect to the nucleon, $\gamma^0$ and $\gamma^5$ refer to the electron. The nucleon $N$ can be either the proton or the neutron.
In the non-relativistic limit this interaction is proportional to the scalar product $(\boldsymbol{\nabla} a, \boldsymbol{s})$, where $\boldsymbol{s}$ is the electron spin \cite{stadnik2015nuclear}. Therefore, this interaction is usually described as the interaction of the axion with the electron spin. The opposite case of the interaction of the axion with the nucleon spin is also possible, it is characterized by product $g_N^p g_e^s$. However, this contribution is suppressed by the factor $1/M$, where $M$ is the nucleon mass \cite{dzuba2018new}.

The axion mass determines the characteristic range of the Yukawa-type interaction,
$$R_{\textrm{Yu}} = 
\frac{1}{m_a} (\textrm{relativistic units}) = 
\frac{\hbar}{m_a c}.$$
This range can be compared with the atomic characteristic of the same dimension $a_B$, Bohr radius. The condition $R_{\textrm{Yu}} = a_B$ implies $m_a = \alpha m_e \approx 4$keV, where $\alpha$ is the fine structure constant. For $m_a \gg \alpha m_e$ 
the effect is determined by the  electronic
density in the narrow vicinity of the nucleus, 
and for $m_a \ll \alpha m_e$  the induced  atomic electric dipole moment (see below) is provided by the whole spatial extent of the electronic wave function.

In the case of one atom, the total potential of this interaction can be expressed as a double sum,
\begin{equation} \label{totalPotential}
    V_{tot}^{(eN)} = \sum_{k=1}^{N_e}\sum_{N} V_{eN}(\boldsymbol{r}_k) = 
    A \bar{g}_N^s g_e^p \sum\limits_{k=1}^{N_e}
    i \frac{e^{-m_a r_k}}{4 \pi r_k} \gamma^0_k \gamma^5_k.
\end{equation}
Here index $N$ runs over all the nucleons of the atomic nucleus, index $k$ enumerates the electrons, the lower index of $\gamma$ matrices determines the electron, which they act on, and $N_e$ is total number of electrons. $A$ is number of nucleons and $\bar{g}_N^s$ is obtained as a parameter $g_N^s$ averaged over nucleons,
\begin{equation}
    \bar{g}_N^s = \frac{Zg_p^s+(A-Z) g_n^s}{A},
\end{equation}
where $Z$ is the atomic nucleus charge, $g_p^s$ and $g_n^s$ are corresponding scalar coupling constants for the proton and neutron, respectively.

The $\mathcal{T}$,$\mathcal{P}$-violating electron-electron ALP-mediated interaction is described by the following two-electron operator \cite{moody1984new,Stadnik:2018}:
\begin{equation}
    V_{ee}(\boldsymbol{r}_1, \boldsymbol{r}_2) = i 
    \frac{g_e^s g_e^p}{4\pi} \frac{e^{-m_a |\boldsymbol{r}_1- \boldsymbol{r}_2|}}{|\boldsymbol{r}_1- \boldsymbol{r}_2|} \gamma^0 \gamma^5.
\end{equation} 
Here $g_e^s$ is the scalar axion-electron coupling constant and $\gamma$-matrices refer to the first electron. For a many-electron system the full axion-mediated electron-electron interaction is given by the Hamiltonian,
\begin{equation} \label{intHam_ee}
    V_{tot}^{(ee)} = \sum\limits_{\substack{i,j=1 \\ i\not=j}}^{N} V_{ee}(\boldsymbol{r}_i, \boldsymbol{r}_j).
\end{equation}

Both the $V_{tot}^{(eN)}$ and the $V_{tot}^{(ee)}$ interactions can induce the $\mathcal{T}$,$\mathcal{P}$-odd permanent atomic electric dipole moment ~\cite{Stadnik:2018}.
If one considers any of them as a perturbation, in the first order of perturbation theory with respect to this interaction, the atomic EDM can be expressed as 
\begin{equation} \label{atomicEDM}
    \boldsymbol{d}_a = \sum\limits_{j>0} 
    \frac{
        \langle \Psi_0| \sum\limits_{i=1}^{N_e} e \boldsymbol{r}_i| \Psi_j \rangle
        \langle \Psi_j | \hat{V}_{tot} | \Psi_0 \rangle
    }{E_0 - E_j} 
    + h.c.
\end{equation}
Here $V_{tot}$ can be either $V_{tot}^{(eN)}$ or $V_{tot}^{(ee)}$, $\boldsymbol{r}_i$ is the radius vector of $i$th electron with respect to the nucleus and $N_e$ is the total electron number. $\Psi_j$ is the electronic wave function of the atom in the $j$th state, $E_j$ is its energy; $j=0$ corresponds to the ground electronic state. 

The induced atomic EDM can be measured as a linear Stark shift in the external electric field. 
Specifically, given that $\boldsymbol{\varepsilon}_{ext}$ is the electric field, the energy shift of atomic level can be calculated as 
\begin{equation}
    \delta E = -d_a \varepsilon_{ext}.
\end{equation}
According to \Eref{atomicEDM}, the value of $d_a$ is proportional to the product $\bar{g}_N^s g_e^p$ for the electron-nucleon interaction or $g_e^s g_e^p$ for the electron-electron one. Therefore, it is possible to introduce electronic structure enhancement factors $\RaxeN$ and $\Raxee$, which depend on $m_a$ and are defined by the following equations:
\begin{equation}
\label{da_eN}
    d_a = \bar{g}_N^s g_e^p \RaxeN ,
\end{equation}
\begin{equation}
\label{da_ee}
    d_a = g_e^s g_e^p \Raxee.
\end{equation}
These coefficients allow one to extract corresponding coupling constants product, when the atomic EDM value (constraint) is obtained experimentally. These coefficients cannot be measured and have to be calculated  by the methods of many-body theory. Their calculation can be performed by using \Eref{atomicEDM} directly, e.g. within the so-called sum over states method. In the present paper we exploit another technique, which is known as the finite-field approach \cite{Monkhorst:77}. It is implemented by calculation of $\RaxeN$ or $\Raxee$ as the following mixed derivative:
\begin{equation} \label{derivative_1e}
\left.    \RaxeN = -\frac{\partial^2 E}{\partial \varepsilon_{ext} \partial (\bar{g}_N^s g_e^p)}\right|_{\substack{\varepsilon_{ext}=0 \\ \bar{g}_N^s g_e^p = 0}},
\end{equation}

\begin{equation} \label{derivative_2e}
\left.    \Raxee = -\frac{\partial^2 E}{\partial \varepsilon_{ext} \partial (g_e^s g_e^p)}\right|_{\substack{\varepsilon_{ext}=0 \\ g_e^s g_e^p = 0}}.
\end{equation}
Its advantage is that calculation of the infinite sum, such as in \Eref{atomicEDM} is not required, but the Schr\"odinger equation should be solved for an atom in an electric field. It requires to reduce the spatial symmetry from $O(3)$ to $C_{\infty v}$ \cite{LL77}.

The many-electron problem has been solved in the following steps. At first, the system of Dirac-Hartree-Fock (DHF)  equations for the many-electron system was solved. Next, the electronic correlation contribution has been calculated. DHF equations have been solved in the basis of Gaussian functions. Specifically, the components of electronic bispinors were expressed as linear combinations of functions $\chi (\boldsymbol{r}) = x^n y^m z^k e^{-\beta r^2}$. Here $\beta >0 $ is the parameter of the exponent, which determines the width of the function, and $n,m,k$ are nonnegative integers; $n+m+k = l$ is the angular momentum of the basis function. 
The primitive one- and two-electron integrals of the Yukawa-type potential over the Gaussian functions for the one-center problem
can be reduced to the following one-dimensional radial integral:
\begin{multline} \label{radialIntegral}
    \int\limits_{0}^{+\infty} dr \ r^{2N-1} e^{-\alpha r^2 - m_a r} =
    \\
    \frac{1}{2\alpha^N}
    \left[
    \Gamma(N) \ _1F_1 \left(N;\frac{1}{2};\frac{m_a^2}{4\alpha}\right)
    \right.
    -
    \\
    -
    \left.
    \frac{m_a}{\sqrt{\alpha}} \cdot \Gamma \left(N+\frac{1}{2}\right)  \ _1F_1 \left(N+\frac{1}{2};\frac{3}{2};\frac{m_a^2}{4\alpha} \right) 
    \right],
\end{multline}
where $\Gamma$ is the Euler $\Gamma$ function, $_1 F _1$ is the Kummer confluent hypergeometric function, and $N$ is a positive integer.

Study of the atomic electronic structure was performed within the coupled cluster approach  \cite{Bartlett1991, Crawford:00, Eliav:1996, eliav:1998, sur:2008}. It is based on the exponential ansatz of the wave function:
\begin{equation} \label{expAnsatz}
    \Psi = \textrm{exp}(\hat{T}) \Phi .
\end{equation}
Here $\Phi$ is the Dirac-Hartree-Fock wave function of the many-electron system and $\Psi$ is the correlated wave function beyond the Dirac-Hartree-Fock approximation. $\hat{T}$ is the cluster excitation operator, which can be written as the following sum:
\begin{equation}
\label{CCexpansion}
    \hat T = \hat{T}_1 + \hat{T}_2 + \hat{T}_3 + \dots,
\end{equation}
where the excitation operators of different orders are defined as follows:
$$
    \hat{T}_1 = \sum\limits_{\substack{i \in occ \\ b \in virt}} t_{i}^{b} a_{b}^\dagger a_{i};
    \quad
    \hat{T}_2 =\frac{1}{2!} \sum\limits_{\substack{i_1<i_2 \in occ \\ b_1<b_2 \in virt}} t_{i_1 i_2}^{b_1 b_2} 
    a_{b_1}^\dagger a_{b_2}^\dagger a_{i_2} a_{i_1}.
$$
Here $t_{\dots}^{\dots}$ are unknown scalar cluster amplitudes to be determined, $a_b^\dagger$ and $a_i$ are creation and annihilation operators of states $b$ and $i$, respectively. Indices $b$ and $i$ correspond to the virtual and occupied states, respectively.  Since the solution of the coupled cluster equations for all the possible excitations for many-electron atoms is unfeasible, the sum (\ref{CCexpansion}) should be truncated. Below we exploit the following commonly used implementations: the coupled cluster approach with single and double cluster amplitudes (CCSD) and the couple cluster approach with single, double, and perturbative triple amplitudes [CCSD(T)]. The former one is based on approximation $\hat{T} \approx \hat{T}_1 + \hat{T}_2$, and the latter one includes also perturbative consideration of the $\hat{T}_3$ term. The CCSD(T) method is often called ``the golden standard'' of quantum chemistry and usually allows one to calculate electronic properties with high accuracy \cite{Visscher:96a,Skripnikov:15a}. 

In the electronic structure calculations the Gaussian nuclear charge distribution model~\cite{Visscher:1997} was used. Following Ref.~\cite{Stadnik:2018}, we did not consider the effect of the finite distributions of nucleons in the operator (\ref{totalPotential}). 
For a related problem where the finite distribution of nucleons was studied, see Refs.~\cite{sahoo2017improved,gharibnejad2015dark}.

The DHF equations have been solved using the {\sc dirac15} code \cite{DIRAC15}. Coupled cluster calculation of the $\RaxeN$ parameter has been also performed  using the {\sc dirac15}, whereas in the case of $\Raxee$  the {\sc mrcc} code \cite{MRCC2020} has been used. The program to construct the matrix of the operator (\ref{totalPotential}) was developed in Ref. \cite{Maison:2021}. The code to construct the matrix of the operator (\ref{intHam_ee}) was developed in Ref.~\cite{maison2021axion} for $m_a \ll 1 \textrm{keV}$ and generalized for arbitrary axion masses in the present paper.

\section{Results and discussion}
We report the results for $m_a = 10^n$eV for $n=$~1,~2,~$\dots 10$. As it was shown in Refs. \cite{Stadnik:2018, Maison:2021}, this range is also sufficient to estimate $\RaxeN(m_a)$ and $\Raxee(m_a)$ functions for extremely high and extremely low values of $m_a$ beyond the considered interval.

For all the values of axion mass, the DHF calculations have been performed within three different  uncontracted Dyall's basis sets of increasing quality: CV2Z, CV3Z and CV4Z \cite{dyall2009relativistic}. All electrons were included in the coupled cluster calculation. The virtual orbitals cutoff threshold was set equal to $10000~E_h$. For comparison, the orbital energy of $1s$ electron is $\sim -3700~E_h$. The importance of high energy orbitals contribution to similar one-electron properties was shown in Refs. \cite{Skripnikov:17a, Skripnikov:15a}. The interaction with the external electric field has been added to the electronic Hamiltonian from the beginning, i.e., already at the DHF level. This corresponds to the ``strategy I'' of Ref.~\cite{Skripnikov:17a}.
The results for the $\RaxeN$ constant for various $m_a$ values within the DHF approach are given in \tref{resultsDF}. Tables \ref{resultsCCSD} and \ref{resultsCCSD_T} include analogous results obtained within the CCSD and CCSD(T) methods, respectively. The final results are highlighted in bold in Table~\ref{resultsCCSD_T}. 
Tables IV--VI contain the results for $\Raxee$ within DHF, CCSD, and CCSD(T) approaches, respectively, using the CV2Z and CV3Z basis sets. The final results are also highlighted in bold in Table VI.

\subsection{Contribtuion of $\RaxeN$}

\begin{table}[h!]
\caption{The values of the $\RaxeN$ parameter for the $^{210}_{\ 87}$Fr atom obtained within the Dirac-Hartree-Fock approach within different basis sets. The results are presented in units of $\textrm{cm}/|e|$.}
\label{resultsDF}
    \begin{tabular}{cccc} 
        \hline
        \hline
        $m_a$ (eV) &$\ $ CV2Z $ \ $ & CV3Z $ \ $& CV4Z $ \ $ \\
        \hline
        $10$  
        & $+1.06 \times 10^{-9}$
        & $+1.06 \times 10^{-9}$
        & $+1.06 \times 10^{-9}$
        \\
        $10^2$  
        & $+1.05 \times 10^{-9}$
        & $+1.05 \times 10^{-9}$
        & $+1.05 \times 10^{-9}$
        \\
        $10^3$  
        & $+7.46 \times 10^{-10}$
        & $+7.46 \times 10^{-10}$
        & $+7.46 \times 10^{-10}$
        \\
        $10^4$  
        & $+1.03 \times 10^{-10}$
        & $+1.02 \times 10^{-10}$
        & $+1.02 \times 10^{-10}$
        \\
        $10^5$  
        & $-3.54 \times 10^{-10}$
        & $-3.54 \times 10^{-10}$
        & $-3.54 \times 10^{-10}$
        \\
        $10^6$  
        & $-4.67 \times 10^{-10}$
        & $-4.67 \times 10^{-10}$
        & $-4.67 \times 10^{-10}$
        \\
        $10^7$  
        & $-3.41 \times 10^{-11}$
        & $-3.41 \times 10^{-11}$
        & $-3.41 \times 10^{-11}$
        \\
        $10^8$  
        & $-8.08 \times 10^{-13}$
        & $-8.18 \times 10^{-13}$
        & $-8.19 \times 10^{-13}$
        \\
        $10^9$  
        & $-9.11 \times 10^{-15}$
        & $-9.35 \times 10^{-15}$
        & $-9.39 \times 10^{-15}$
        \\
        $10^{10}$  
        & $-9.13 \times 10^{-17}$
        & $-9.38 \times 10^{-17}$
        & $-9.41 \times 10^{-17}$
        \\ 
    \hline
    \hline
    \end{tabular}
\end{table}

\begin{table}[h!]
\caption{Results of $\RaxeN$ parameter calculations for various axion masses within the CCSD approach in units of cm/$|e|$.}
\label{resultsCCSD}
    \begin{tabular}{cccc}
        \hline
        \hline
        $m_a$ (eV) &$\ $ CV2Z $ \ $ & CV3Z $ \ $& CV4Z $ \ $ \\
        \hline
        $10$  
        & $+1.09 \times 10^{-9}$
        & $+1.10 \times 10^{-9}$
        & $+1.11 \times 10^{-9}$
        \\
        $10^2$  
        & $+1.08 \times 10^{-9}$
        & $+1.10 \times 10^{-9}$
        & $+1.10 \times 10^{-9}$
        \\
        $10^3$  
        & $+8.24 \times 10^{-10}$
        & $+8.40 \times 10^{-10}$
        & $+8.43 \times 10^{-10}$
        \\
        $10^4$  
        & $+1.24 \times 10^{-10}$
        & $+1.28 \times 10^{-10}$
        & $+1.29 \times 10^{-10}$
        \\
        $10^5$  
        & $-4.58 \times 10^{-10}$
        & $-4.69 \times 10^{-10}$
        & $-4.71 \times 10^{-10}$
        \\
        $10^6$  
        & $-5.71 \times 10^{-10}$
        & $-5.88 \times 10^{-10}$
        & $-5.91 \times 10^{-10}$
        \\
        $10^7$  
        & $-4.17 \times 10^{-11}$
        & $-4.29 \times 10^{-11}$
        & $-4.32 \times 10^{-11}$
        \\
        $10^8$  
        & $-9.88 \times 10^{-13}$
        & $-1.03 \times 10^{-12}$
        & $-1.04 \times 10^{-12}$
        \\
        $10^9$  
        & $-1.11 \times 10^{-14}$
        & $-1.18 \times 10^{-14}$
        & $-1.19 \times 10^{-14}$
        \\
        $10^{10}$  
        & $-1.12 \times 10^{-16}$
        & $-1.18 \times 10^{-16}$
        & $-1.19 \times 10^{-16}$
        \\
    \hline
    \hline
    \end{tabular}
\end{table}

\begin{table}[h!]
\caption{Results of $\RaxeN$ parameter calculations for various axion masses within the CCSD(T) approach in units of cm/$|e|$.}
\label{resultsCCSD_T}
    \begin{tabular}{cccc}
        \hline
        \hline
        $m_a$ (eV) &$\ $ CV2Z $ \ $ & CV3Z $ \ $& CV4Z $ \ $ \\
        \hline
        $10$  
        & $+1.09 \times 10^{-9}$
        & $+1.10 \times 10^{-9}$
        & $\boldsymbol{+1.10 \times 10^{-9}}$
        \\
        $10^2$  
        & $+1.09 \times 10^{-9}$
        & $+1.09 \times 10^{-9}$
        & $\boldsymbol{+1.09 \times 10^{-9}}$
        \\
        $10^3$  
        & $+8.28 \times 10^{-10}$
        & $+8.38 \times 10^{-10}$
        & $\boldsymbol{+8.39 \times 10^{-10}}$
        \\
        $10^4$  
        & $+1.24 \times 10^{-10}$
        & $+1.28 \times 10^{-10}$
        & $\boldsymbol{+1.28 \times 10^{-10}}$
        \\
        $10^5$  
        & $-4.59 \times 10^{-10}$
        & $-4.67 \times 10^{-10}$
        & $\boldsymbol{-4.68 \times 10^{-10}}$
        \\
        $10^6$  
        & $-5.73 \times 10^{-10}$
        & $-5.86 \times 10^{-10}$
        & $\boldsymbol{-5.89 \times 10^{-10}}$
        \\
        $10^7$  
        & $-4.18 \times 10^{-11}$
        & $-4.28 \times 10^{-11}$
        & $\boldsymbol{-4.30 \times 10^{-11}}$
        \\
        $10^8$  
        & $-9.91 \times 10^{-13}$
        & $-1.03 \times 10^{-12}$
        & $\boldsymbol{-1.03 \times 10^{-12}}$
        \\
        $10^9$  
        & $-1.12 \times 10^{-14}$
        & $-1.17 \times 10^{-14}$
        & $\boldsymbol{-1.18 \times 10^{-14}}$
        \\
        $10^{10}$  
        & $-1.12 \times 10^{-16}$
        & $-1.18 \times 10^{-16}$
        & $\boldsymbol{-1.19 \times 10^{-16}}$
        \\
    \hline
    \hline
    \end{tabular}
\end{table}

Analysis of the results in Table III demonstrates that the $\RaxeN$ value changes its sign in the interval of 
$\left(10^4\ \textrm{eV}; 10^5\ \textrm{eV}\right)$. The reason is that for low-mass axions $\RaxeN$ is determined by the oscillating wave function of the valence electron in a wide range, whereas for high $m_a$ only the close vicinity of the nucleus gives a noticeable contribution.

For low-mass axions ($m_a \ll 1$keV) the $\RaxeN$ parameter is weakly dependent on both the basis set size and the electronic correlation treatment level. In contrast, for the heavy axions, the role of correlation effects is sufficiently higher: for $m_a = 10^{10}$~eV contribution of these effects reaches 20\%, whereas for $m_a = 10$~eV it is less than 4\%. Additionally, for low-mass axions $\RaxeN$ is almost independent on $m_a$, whereas for large $m_a$ the dependence is significant.  For $m_a \gg 1$~MeV the inverse quadratic dependence takes place: $\RaxeN \sim m_a^{-2}$. 
Therefore, it is possible to introduce parameter $\tilde{R}$ in the following way:
\begin{equation}
    \tilde{R} =  \lim\limits_{m_a \rightarrow \infty} m_a^2\RaxeN(m_a).
\end{equation} It will be used further in order to estimate the atomic EDM in the heavy axion case.

In the case of high $m_a$, $\RaxeN$ can be estimated in another way. For this, one can use the similarity of \Eref{intHam_eN} and the Hamiltonian of the scalar-pseudoscalar nucleus-electron interaction in the point-like nucleus approximation \cite{Ginges:04}:
\begin{equation} \label{SPS}
    H_{T,P} = k_{T,P} \cdot iZ \frac{G_F}{\sqrt{2}} \sum\limits_k \gamma^0_k \gamma^5_k \delta(\boldsymbol{r}_k).
\end{equation}
Here $k_{T,P}$ is the parameter of the $\mathcal{T}$,$\mathcal{P}$-odd  interaction and $G_F \approx 1.166 \cdot 10^{-23}\  \textrm{eV}^{-2} (\hbar c)^3$ is the Fermi constant. Taking also into account the relation,
\begin{equation} \label{weakLimit}
    \frac{e^{-m_a r}}{4 \pi r}  \approx  \frac{1}{m_a^2}  \delta(\boldsymbol{r}),
\end{equation}
one obtains for large $m_a$,
\begin{equation}
    \RaxeN(m_a) \approx \frac{A \sqrt{2}}{Z} \cdot \frac{1}{m_a^2 G_F} R_s,
\end{equation}
where the atomic enhancement parameter of the scalar-pseudoscalar nucleus-electron interaction $R_s$ is defined similarly to \Eref{derivative_1e}. One can see from Tables I--III, that the inverse quadratic dependence with respect to the axion mass is fulfilled with high accuracy for $m_a \gg 1~\textrm{MeV}$.

Obtained results  for $\RaxeN$ can be compared with the ones for other atoms. For example, according to Ref. \cite{Stadnik:2018} on the borders of the interval studied:
$$ \RaxeN \left( _{\ 55}^{133}\textrm{Cs}, m_a = 10~\textrm{eV} \right) \approx 
+7.3 \times 10^{-10}~\textrm{cm}/|e|,  $$
$$ \RaxeN \left( _{\ 55}^{133}\textrm{Cs}, m_a = 10^{10}~\textrm{eV} \right) \approx 
-6.8 \times 10^{-18}~\textrm{cm}/|e|. $$
For low-mass axions, the ratio of $\RaxeN$ constants for $_{\ 87}^{210}$Fr and $_{\ 55}^{133}$Cs is close to
$Z_{\textrm{Fr}}/Z_{\textrm{Cs}} \approx A_{\textrm{Fr}}/A_{\textrm{Cs}} \approx 1.6$. 
However, in the heavy axion case, the $\RaxeN$ value increases dramatically when the nuclear charge increases. The character of this dependence is extensively analyzed in Ref. \cite{Stadnik:2018}.

Provided results allow one to estimate the atomic EDM $d_a$ induced by the studied interaction. According to the updated results of Ref.~\cite{Stadnik:2018}, the upper constraint for coupling constants product 
was obtained in the ThO beam experiment \cite{ACME:18} 
\footnote{The best constraints on other products of axion-fermion coupling constants considered below also follow from the ThO experiment \cite{ACME:18,Stadnik:2018}.}
and is $|\bar{g}_N^s g_e^p| \lesssim~9.0 \times 10^{-20}~\hbar c$ under the condition $m_a \ll 1 \textrm{keV}$. According to 
\Eref{da_eN},
this corresponds to the atomic EDM constraint
$|d_a(\textrm{Fr})| \lesssim 1.4 \times 10^{-26} |e|\ \textrm{cm}$.
Now, let us consider the high-mass axion limit.
Substitution of the constraint value $|\bar{g}_N^s g_e^p|/m_a^2 \lesssim 6.0 \times 10^{-15}~\hbar c \  \textrm{GeV}^{-2}$ from Ref.~\cite{Stadnik:2018} and result of the present study $\tilde{R} = -1.19 \times 10^{-14}~\textrm{GeV}^2 \textrm{cm}/|e|$ shows that the atomic EDM induced by heavy axion exchange between nucleon and electron correspond to the atomic EDM value 
$|d_a(\textrm{Fr})| \lesssim 9.8 \times 10^{-27} |e|\ \textrm{cm}$. 
These constraints can be compared with Fr atomic EDM, induced by the $e$EDM, 
\begin{equation}
    |d_a (e\textrm{EDM})| = |d_e R_d| \lesssim 10^{-26} |e|\ \textrm{cm},
\label{deFrEDM}    
\end{equation} 
where $|d_e| \leq 1.1 \cdot 10^{-29} |e| \ \textrm{cm}$ is the current constraint on the $e$EDM \cite{ACME:18} and $R_d \approx 800$ is the dimensionless $e$EDM enhancement factor~\cite{Byrnes:99, Mukherjee:09, Blundell:2012, Roberts:13, shitara2021cp}. 
The axion-induced EDM values in both considered cases have the same order of magnitude as the upper limit on the $e$EDM contribution. Therefore, if the constraint on the Fr EDM, better than in Eq.~(\ref{deFrEDM}) is experimentally obtained, it will also lead to a new constraint on the product of the considered coupling constants.

\subsection{Contribution of $\Raxee$}

\begin{table}[h!]
\caption{Results of calculations of the $\Raxee$ parameter for the $^{210}_{\ 87}$Fr atom within the DHF-approach in units of cm/$|e|$.}
\label{results_HF_2e}
    \begin{tabular}{ccc}
    \hline
    \hline
    $m_a$ (eV) & CV2Z & CV3Z \\
    \hline
    $10$   & $+4.37\times 10^{-10}$& $+4.37\times 10^{-10}$ \\
    $10^2$ & $+4.34\times 10^{-10}$& $+4.34\times 10^{-10}$ \\
    $10^3$ & $+3.10\times 10^{-10}$& $+3.10\times 10^{-10}$\\
    $10^4$ & $+7.21\times 10^{-11}$& $+7.21\times10^{-11}$\\
    $10^5$ & $+9.34\times 10^{-12}$& $+9.33\times10^{-12}$ \\
    $10^6$ & $-1.35\times 10^{-13}$&$-1.35\times10^{-13}$\\
    $10^7$ & $-7.13\times 10^{-15}$& $-7.13 \times10^{-15}$\\
    $10^8$ & $-7.37\times 10^{-17}$& $-7.37\times10^{-17}$\\
    $10^9$ & $-7.38\times10^{-19}$&$-7.37\times10^{-19}$ \\
    $10^{10}$ &$-7.38\times10^{-21}$ & $-7.37\times10^{-21}$\\
    \hline
    \hline
    \end{tabular}
\end{table}

\begin{table}[h!]
\caption{Results of calculations of the $\Raxee$ parameter for the $^{210}_{\ 87}$Fr atom within the CCSD-approach in units of cm/$|e|$.}
\label{results_CCSD_2e}
    \begin{tabular}{ccc}
    \hline
    \hline
    $m_a$ (eV) & CV2Z & CV3Z \\
    \hline
    $10$ & $+4.52 \times 10^{-10}$& $+4.57 \times 10^{-10}$ \\
    $10^2$ & $+4.49\times10^{-10}$& $+4.54\times 10^{-10}$ \\
    $10^3$ & $+3.46\times10^{-10}$& $+3.52\times 10^{-10}$\\
    $10^4$ & $+9.31\times10^{-11}$& $+9.59\times10^{-11}$\\
    $10^5$ & $+1.13\times10^{-11}$& $+1.16\times10^{-11}$ \\
    $10^6$ & $-1.82\times10^{-13}$&$-1.86\times10^{-13}$\\
    $10^7$ & $-8.83 \times10^{-15}$& $-9.07 \times10^{-15}$\\
    $10^8$ & $-9.13\times10^{-17}$& $-9.38\times10^{-17}$\\
    $10^9$ & $-9.14\times10^{-19}$&$-9.39\times10^{-19}$ \\
    $10^{10}$ &$-9.14\times10^{-21}$ & $-9.39\times10^{-21}$\\
    \hline
    \hline
    \end{tabular}
\end{table}

\begin{table}[h!]
\caption{Results of calculations of the $\Raxee$ parameter for the $^{210}_{\ 87}$Fr atom within the CCSD(T)-approach in units of cm/$|e|$.}
\label{results_CCSD(T)_2e}
    \begin{tabular}{ccc}
    \hline
    \hline
    $m_a$ (eV) & CV2Z & CV3Z \\
    \hline
    $10$ & $+4.40 \times 10^{-10}$& $\boldsymbol{+4.38 \times 10^{-10}}$ \\
    $10^2$ & $+4.37\times10^{-10}$& 
    $\boldsymbol{+4.36\times 10^{-10}}$ \\
    $10^3$ & $+3.37\times10^{-10}$ & 
    $\boldsymbol{+3.38\times 10^{-10}}$\\
    $10^4$ & $+9.09\times10^{-11}$& $\boldsymbol{+9.25\times10^{-11}}$\\
    $10^5$ & $+1.10\times10^{-11}$& $\boldsymbol{+1.12\times10^{-11}}$ \\
    $10^6$ & $-1.77\times10^{-13}$ &
    $\boldsymbol{-1.79\times10^{-13}}$\\
    $10^7$ & $-8.61 \times10^{-15}$ & 
    $\boldsymbol{-8.75 \times10^{-15}}$\\
    $10^8$ & $-8.90\times10^{-17}$ & $\boldsymbol{-9.05\times10^{-17}}$\\
    $10^9$ & $-8.90\times10^{-19}$ & $\boldsymbol{-9.05\times10^{-19}}$\\
    $10^{10}$ &$-8.91\times10^{-21}$ & $\boldsymbol{-9.05\times10^{-21}}$\\
    \hline
    \hline
    \end{tabular}
\end{table}

The $\Raxee$ parameter changes its sign for the $m_a$ value which is higher than the value for the $\RaxeN$ case (see above) -- in the interval $\left(10^5\ \textrm{eV},\ 10^6\ \textrm{eV}\right)$. As in the case of the one-electron interaction, results of the DHF approach are weakly dependent on the basis set size. For low values of $m_a$ results of DHF and CCSD(T) approaches almost coincide, whereas CCSD results slightly differ from them.

Comparison of the final results shows one that the ratio $\RaxeN / \Raxee$ is close to $A/Z$ for $m_a \ll 1 \textrm{keV}$. This result is consistent with the same proportion for the YbOH molecule \cite{Maison:2021, maison2021axion} and heavy atoms considered in Ref.~\cite{Stadnik:2018}. 

Results of Ref. \cite{Stadnik:2018} for the caesium atom are as follows:
$$\Raxee  \left( _{\ 55}^{133}\textrm{Cs}, m_a = 10~\textrm{eV} \right) \approx +2.9 \times 10^{-10}  ~\textrm{cm}/|e|,$$
$$\Raxee  \left( _{\ 55}^{133}\textrm{Cs}, m_a = 10^{10}~\textrm{eV} \right) \approx -2.8 \times 10^{-21}  ~\textrm{cm}/|e|.$$ 
As for the electron-nucleon interaction case, for small $m_a$ the ratio $\Raxee(\textrm{Fr}) / \Raxee(\textrm{Cs}) \approx Z_{\textrm{Fr}}/ Z_{\textrm{Cs}}$ is fulfilled. However, for heavy axions, the proportion is $\Raxee(\textrm{Fr})/\Raxee(\textrm{Cs}) \sim \left(Z_{\textrm{Fr}}/Z_{\textrm{Cs}}\right)^2$, which is consistent with Eq. (18) of Ref. \cite{Stadnik:2018}.

Despite the fact that Eq.~(\ref{weakLimit}) is not valid in the two-particle case, $\Raxee(m_a)$ is also proportional to $m_a^{-2}$. This can be explained by the fact that the main contribution to $\Raxee$ is given by the mixing of $s$  and $p$ electrons, and the corresponding matrix elements are proportional to expression (\ref{radialIntegral}) with $N=1$; and in this case the integral is scaled as $m_a^{-2}$.

Combining the final results for $\Raxee$ for extremely low and extremely high mass with the constraints from Ref.  \cite{Stadnik:2018} ($|g_e^s g_e^p| \lesssim 2.4\times 10^{-19} \hbar c$ and $|g_e^s g_e^p/m_a^2 \lesssim 5.6 \times 10^{-11} \hbar c \ \textrm{GeV}^{-2}$, respectively), one can obtain the upper limit of the electron-electron axion-induced contribution to the atomic EDM:
\begin{equation}
    |d_a^{(ee)}(m_a \ll 1\textrm{keV})| \lesssim 1.4 \times 10^{-26}|e| \ \textrm{cm},
\end{equation}
\begin{equation}
    |d_a^{(ee)}(m_a \gg 1\textrm{MeV})| \lesssim 7.0 \times 10^{-27}|e| \ \textrm{cm}.
\end{equation}
These values are of the same order of magnitude as for the electron-nucleon interaction case. So, the constraint on $g_e^s g_e^p$ may also be updated if the sensitivity, better than in Eq.~(\ref{deFrEDM}), will be obtained in the francium atomic EDM measurement.

\begin{acknowledgments}
    D.E.M. is grateful to M.Y. Reznikov for his assistance in the two-electron problem solution. Electronic structure calculations have been carried out using computing resources of the federal collective usage center Complex for Simulation and Data Processing for Mega-science Facilities at National Research Centre ``Kurchatov Institute'', http://ckp.nrcki.ru/.

    $~~~$Molecular coupled cluster electronic structure calculations have been supported by the Russian Science Foundation Grant No. 19-72-10019. Calculations of the $\RaxeN$ matrix elements were supported by the Foundation for the Advancement of Theoretical Physics and Mathematics ``BASIS'' Grant according to Projects No. 20-1-5-76-1 and No. 21-1-2-47-1.
\end{acknowledgments}

\bibliographystyle{apsrev}

\end{document}